\documentclass[11pt,a4paper,twoside,groupcitations]{article}
\usepackage[T1]{fontenc}
\usepackage[ansinew]{inputenc}
\usepackage[english]{babel}
\usepackage{amsfonts}
\usepackage{amsmath}
\usepackage{array}
\usepackage{amsthm}
\usepackage{amssymb}
\usepackage{graphicx}
\usepackage{subfigure}
\usepackage{braket}
\usepackage{eucal}
\usepackage{verbatim}
\usepackage[table]{xcolor}
\usepackage{caption}
\usepackage{bm}
\usepackage{multicol}
\usepackage{tikz}
\usetikzlibrary{positioning,arrows}
\usetikzlibrary{decorations.pathmorphing}
\usetikzlibrary{decorations.markings}
\usetikzlibrary{calc,decorations.markings}
\usetikzlibrary{arrows,shapes}
\usetikzlibrary{matrix,arrows}
\usepackage{pgfplots}
\usepackage{xparse}
\definecolor{jade}{HTML}{00A86B}
\newcommand{\be}{\begin{eqnarray}}
\newcommand{\ee}{\end{eqnarray}}

\renewcommand{\d}{\mbox{${\rm d}$}} 
\newcommand{\lp}{\ell_{\rm p}}
\newcommand{\mpl}{m_{\rm p}}

\newcommand{\bg}{\bar g}
\newcommand{\babla}{\bar\nabla}
\title{\bf Domestic Corpuscular Inflaton}
\author{Andrea~Giugno$^{a}$\thanks{E-mail: A.Giugno@physik.uni-muenchen.de}
$\ $ and
Andrea~Giusti$^{abc}$\thanks{E-mail: andrea.giusti@bo.infn.it}
\\
\\
$^a${\em Arnold Sommerfeld Center, Ludwig-Maximilians-Universit\"at}
\\
{\em Theresienstra{\ss}e~37, 80333 M\"unchen, Germany}
\\
\\
$^b${\em Dipartimento di Fisica e Astronomia, Universit\`a di Bologna}
\\
{\em via Irnerio~46, 40126 Bologna, Italy}
\\
\\
$^c${\em I.N.F.N., Sezione di Bologna, I.S.~FLAG}
\\
{\em viale B.~Pichat~6/2, 40127 Bologna, Italy}
}
\begin{document}
\maketitle
\begin{abstract}
The aim of this paper is to provide a more precise description of the paradigm of
corpuscular slow-roll inflation, which was previously introduced by Casadio \emph{et al.} 
in~\cite{Casadio:2017twg}. Specifically, we start by expanding the Starobinsky theory
on a curved background and then infer the number and nature of the propagating degrees of
freedom, both in the true inflationary phase and in a quasi-de Sitter approximation. 
We correctly find that the particle spectrum contains a transverse trace-free mode and a scalar one.
The scalar mode displays a tachyonic nature during the slow-roll phase, due to the instability
of the system, whereas it acquires the appropriate oscillatory behavior as the background 
approaches a critical value of the curvature. These results confirm the fact that the Einstein-Hilbert 
term acts as a perturbation to the quadratic one, and is responsible
for driving the early universe out of the inflationary phase, thus realising the inflaton field
in terms of pure (corpuscular) gravity.
\end{abstract}
\section{Introduction}
\setcounter{equation}{0}
\label{Sintro}
One of the most successful accomplishments of cosmology is the description of the evolution of the early Universe,
from a small region of extreme heat and density all the way to large-scale structures. Unfortunately, taking this scenario at face value during its very beginning thwarts the principle of homogeneity at late times. Moreover, it would look very hard to explain why the matter density is so close to the critical value, unless one bestows a colossal amount of fine tuning upon the physics of the early Universe.

These problems are tackled down by the theory of cosmic inflation, which entails a period of exponential expansion shortly after the Big Bang. Following the original proposals of Guth~\cite{Guth-first} and Starobinsky~\cite{starobinskyI}, a new paradigm emerged thanks to Linde~\cite{Linde} and Albrecht and Steinhardt~\cite{AS}. This new approach considers a scalar degree of freedom (the inflaton) rolling down a potential plateau, rather than a vacuum-to-vacuum transition. When the inflaton starts oscillating around the minimum, inflation stops and the radiation dominated phase begins, the whole process going under the name of reheating~\cite{Kofman}.

	The most common inflationary models are semiclassical, as they describe the dynamics of quantum mechanical degrees of freedom on classical background geometries. While this set up may seem sufficient to attempt a solution to the horizon and flatness problems, it misses the contribution of quantum gravitational effects, which may have been relevant in the time frame of the early Universe. However, one can keep this line of thought and suggest that the emergence of the classical background is just the effect of a coherent quantum state of a large number of gravitons, in affinity to the description of a laser beam in terms of a collinear bunch of photons. This point of view is forwarded by the corpuscular model of gravity~\cite{dvali}, according to which the strong coupling regime is achieved by the formation of extended quasi-classical objects, composed by a huge number of soft degrees of freedom. Since the gravitational interaction is purely attractive, the constituents can superimpose in the same quantum state, effectively giving rise to a self-gravitating compact object, roughly shaped as a Bose-Einstein Condensate (BEC). A vast literature~\cite{Gia1} applies this philosophy to black holes, which are thought of as quantum systems on the verge of a phase transition. In particular, the black hole interior is understood as a state of off-shell gravitons, which are confined in an effective potential well of the size of their Compton/de~Broglie wavelength $\lambda$, and whose marginal interaction is regulated by the effective coupling $\alpha_{\rm eff}\sim 1/N$. 

	A couple of interesting results come straightforward. First off, Hawking radiation is understood in terms of 
leakage of the condensate due to mutual scattering of soft gravitons and the famous Bekenstein-Hawking area law~\cite{bekenstein} is promptly retrieved, together with the correct logarithmic corrections~\cite{planckian}. In addition to that, the quantum corrections to the condensate, at the leading order, reproduce the appropriate post-Newtonian expansion of the (weak) gravitational field surrounding a static spherically symmetric source~\cite{Casadio:2017cdv}, once the contribution of collapsing matter is correctly taken into account~\cite{baryons}. 
\par
 Beside its apparent effectiveness in describing objects of stellar size, this approach provides also some interesting new insight when applied to the Universe as a whole~\cite{dvali,cko}. In particular, it is of paramount interest to see whether corpuscular gravity corrects experimentally relevant parameters to an appreciable extent, see e.g.~\cite{CMB,WMAP}.

	It has been pointed out~\cite{Casadio:2017twg} that the self-interaction of the gravitons can embed the inflationary scenario in a corpuscular perspective, and the primordial cosmological condensate can reproduce this machinery without requiring the insertion of an exotic degree
of freedom in the particle spectrum.
The Starobinsky model, implemented as an $f(R)$ theory~\cite{starobinsky,Capozziello:2009nq,Capozziello:2011et,DeLaurentis:2015fea,DeFelice:2010aj} with $f(R)=R+\alpha R^2$, is employed as reference setting for the inflationary scenario. It is then argued that when $R^2$ dominates, the background can be approximated by an ideal de Sitter spacetime, reflecting the scale invariance of the curvature-squared term. The latter is in turn perturbed by the Einstein-Hilbert part of action, namely $R$, and the Universe is driven slowly out of inflation as a result.

	Besides, it is also woth remarking that the cosmological condensate is an interesting tool to test a plethora of modified gravity theories, such as Modified Newtonian Dynamics (MOND)~\cite{Cadoni:2017evg,Cadoni:2018dnd}.

In this paper we clarify some of the statements made in past works about corpuscular inflation. First, we establish what are the degrees of freedom propagating on the approximate de Sitter background, matching known results~\cite{deRham:2012kf,Biswas:2016etb}. Consequently, we show that there is a tachyonic polarization in the spectrum and that our approach agrees with the theory of an eternally inflating space, as one would expect from de Sitter. Then, we show that this degree of freedom is driven out of the tachyonic phase once the Einstein-Hilbert term perturbs the action, explicitly realising what was theorised in~\cite{Casadio:2017twg}. Finally, we present some concluding remarks and possible outlook.
\section{Starobinsky action on curved backgrounds}
\setcounter{equation}{0}
We consider the $f(R)$ theory reproducing the inflationary model of Starobinsky. It is described
in terms of the action
\be
{\rm S}
\,=\,
\frac{\mpl}{16\pi \lp} \int \d^4 x \, \sqrt{-g}\left( R+\alpha R^2 \right)
\label{StaroAction}
\ ,
\qquad
\alpha
\,=\,
\frac{\lp^2\mpl^2}{6M^2}
\ee
where $R$ is the curvature scalar, $M$ a scale mass and we follow the conventions listed in Appendix~\ref{sec:Conv}. 
The variation of $\rm S$ with respect to $g^{\mu\nu}$~\eqref{Variation}
gives the tensorial equations of motion (EOM for short)
\be
R_{\mu\nu}-\frac{1}{2}g_{\mu\nu}R+\alpha R\left(2R_{\mu\nu}-\frac{1}{2}g_{\mu\nu}R \right)
+2\alpha\left(g_{\mu\nu}\Box-\nabla_\mu\nabla_\nu\right) R
\,=\,
0
\ ,
\label{EOM}
\ee
which we rewrite as ${\cal E}_{\mu\nu}=0$ in compact notation.
The first two terms are the Einstein tensor and represent the general relativity (GR) part of the theory, whereas the others contain high order derivatives.

We manipulate this tensorial equation by taking covariant divergence and trace. Thus, the
contraction with $\nabla^\mu$ vanishes identically, since
\be
\nabla^\mu {\cal E}_{\mu\nu}
\,=\,
2\alpha R_{\mu\nu}\nabla^\mu R +2\alpha(\nabla_\nu \Box-\Box \nabla_\nu)R
\equiv
0
\ ,
\label{DivEOM}
\ee
thanks to Eq.~\eqref{CommuteNabla}, the contracted Bianchi identities
$2\nabla^\mu R_{\mu\nu}-\nabla_\nu R=0$ and the commutation
\be
\Box\nabla_\nu R
\,=\,
\nabla_\nu\Box R+R_{\mu\nu}\nabla^\mu R
\ .
\ee
This result is reasonable because otherwise we would single out an EOM for a would-be vectorial
polarization of the graviton, which is known to be non-propagating. 

On a curved space-time, the on-shell graviton possesses one tensorial and one scalar polarization, which are the ones
we aim to specify.
The trace of EOM yields
\be
\left(6\alpha\,\Box-  1\right) R
\,=\,
0
\label{TraceEOM}
\ ,
\ee
which can be used to simplify the EOM~\eqref{EOM},
\be
R_{\mu\nu}-\frac{1}{6}g_{\mu\nu}R+\alpha R\left(2R_{\mu\nu}-\frac{1}{2}g_{\mu\nu}R \right)
-2\alpha\nabla_\mu\nabla_\nu R
\,=\,
0
\ .
\label{EOMreduced}
\ee
\par
So far, we checked a fact well-known in literature~\cite{Capozziello:2011et}: the graviton propagates two kind of
waves on a curved background. Unfortunately, only the full metric appear, and we
have no insight on the behavior of the fluctuations over the background.
\par
Now, we consider the linearized theory~\cite{Straumann}
\begin{subequations}
\be
g_{\mu\nu}
\!\!&=&\!\!
\bg_{\mu\nu}+\epsilon h_{\mu\nu}
\ ,
\\
g^{\mu\nu}
\!\!&=&\!\!
\bg^{\mu\nu}-\epsilon h^{\mu\nu}+\mathcal{O}(\epsilon^2)
\ ,
\ee
\label{var_metr}
\end{subequations}
around a fiducial and purely classical (and curved) metric, $\bg_{\mu\nu}$~\footnote{
The fact that the background is curved means that we do not have
 $\partial_\lambda \bg_{\mu\nu}=0$ in general but $\babla_\lambda \bg_{\mu\nu}=0$.
 As we explain later on, barred quantities are computed on $\bg$.}. 
We expand a tensor $\bm{T}$ as
\be
\bm{T}
\,\rightarrow\,
\overline{\bm{T}} +\epsilon \, \delta \bm{T} +\mathcal{O}(\epsilon^2)
\ ,
\label{shift}
\ee
where barred terms are intended as computed over $\bg_{\mu\nu}$.

	First, we expand the Christoffel symbols~\eqref{Christoffel} according to~\eqref{shift},
\be
\delta \Gamma^\lambda_{\mu\nu}
\,=\,
\frac{1}{2} \, \bg^{\lambda \rho}\left(
\babla_\mu h_{\rho \nu}+\babla_\nu h_{\mu\rho}
-\babla_\rho h_{\mu\nu}
\right)
\ ,
\ee
in order to get the linearization of the Ricci tensor, thanks again to the Palatini identity~\eqref{Palatini},
\be
\delta R_{\mu\nu}
\,=\,
\frac{1}{2}\biggl(
-\bar \Box h_{\mu\nu}+\babla^\lambda \babla_\mu
h_{\lambda\nu}+\babla^\lambda \babla_\nu h_{\mu\lambda} -\babla_\nu\babla_\mu h^\lambda_{\,\,\lambda}
\biggr)
\ ,
\label{deltaRmn}
\ee
whose contraction with $\bg^{\mu\nu}$ 
gives the linearization of the curvature scalar $\delta R$
\be
\delta R
\,=\,
-h^{\mu\nu}\bar R_{\mu\nu}+\babla^\mu\babla^\nu h_{\mu\nu}
-\bar \Box h^\mu_{\,\,\mu}
\ .
\label{deltaR}
\ee
These are nothing but Eqs.~\eqref{Palatini} and~\eqref{deltaRapp}, with the insertion of 
the metric variations~\eqref{var_metr} above.
We are going to use these tools in the next Section, where we will solve appropriate
simpler versions of the EOM~\eqref{EOM} on a specified background $\bg_{\mu\nu}$.
\par
The most important implication for us is that the invariance of the theory under the diffeomeorphism
\be
x^\mu
\,\longrightarrow\,
x'^\mu
\,=\,
x^\mu +\xi^\mu(x)
\ee
entails the gauge condition
\be
h_{\mu\nu}
\,\rightarrow\,
h'_{\mu\nu}
\,=\,
h_{\mu\nu}+\babla_{\mu} \xi_\nu + \babla_\nu \xi_\mu
\ ,
\label{hGauge}
\ee
on a curved background. 
As usual, $\xi^\mu$ is just a set of four parameters and has the dimensions of a length.
However, Eq.~\eqref{hGauge} is not the end of the story, since scalar polarizations
appear in addition to the transverse trace-free and the transverse vectorial ones.
We can guess their embedding by requiring symmetry in the
indices and the employing of not more than two derivatives:
\be
h_{\mu\nu}
\,=\,
h^\bot_{\mu\nu}+\lp\left(\babla_{\mu} A^\bot_\nu + \babla_\nu A^\bot_\mu\right)
+\frac{1}{3} \, \bg_{\mu\nu}\varphi +\frac{\babla_\mu \babla_\nu}{\bar \Box} \chi
\label{decomposition}
\ .
\ee 
The factor $\lp$ is to have all the fields with the same dimensionality.
We can use this decomposition to stick to transverse and traceless components, that is
\be
\bg^{\mu\nu}h^\bot_{\mu\nu}
\,=\,
0
\ ,
\quad
\babla^\mu h^\bot_{\mu\nu}
\,=\,
0
\quad {\rm and} \quad
\babla^\mu A^\bot_\mu
\,=\,
0
\ .
\label{TransvTraceless}
\ee
In addition to that, if we contract $h^\bot_{\mu\nu}$ with the background field equation~\eqref{EOM}
we obtain the useful relation
\be
(1+2\alpha \bar R)\bar R^{\mu\nu}h^\bot_{\mu\nu}
\,=\,
2\alpha h^\bot_{\mu\nu}\babla^\mu\babla^\nu \bar R
\ .
\label{h_barEOM}
\ee
\par
\par
It is possible to see~\cite{Odintsov:1991nd,Donoghue:2018izj} that these fields relate to the decomposition of the graviton
under the rotation subgroup of the full Lorentz group. The following prescriptions are useful
\begin{subequations}
\be
h^\bot_{\mu\nu}
\!\!&=&\!\!
h_{\mu\nu}-\frac{\babla_\mu}{\bar \Box}\left(\bg_{\nu\rho}-\frac{\babla_\nu \babla_\rho}{\bar \Box}\right)\babla_\sigma h^{\rho\sigma}
-\frac{\babla_\nu}{\bar \Box}\left(\bg_{\mu\rho}-\frac{\babla_\mu \babla_\rho}{\bar \Box}\right)\babla_\sigma h^{\rho\sigma}
\notag
\\
&&\,\,
-\frac{1}{3}\left(\bg_{\mu\nu}-\frac{\babla_\mu \babla_\nu}{\bar \Box}\right)
\left(\bg_{\rho\sigma}-\frac{\babla_\rho \babla_\sigma}{\bar \Box}\right)h^{\rho\sigma}
-\frac{\babla_\mu\babla_\nu}{\bar \Box}\frac{\babla_\rho \babla_\sigma}{\bar \Box}h^{\rho\sigma}
\ ,
\\
A^\bot_\mu
\!\!&=&\!\!
\frac{1}{\lp\bar \Box} 
\left(\bg_{\mu\nu}-\frac{\babla_\mu\babla_\nu}{\bar \Box}\right)\babla_\rho h^{\rho\nu}
\ ,
\\
\varphi
\!\!&=&\!\!
\left(\bg_{\mu\nu}-\frac{\babla_\mu\babla_\nu}{\bar \Box}\right) h^{\mu\nu}
\ ,
\\
\chi
\!\!&=&\!\!
\frac{1}{3}\left(4\frac{\babla_\mu\babla_\nu}{\bar \Box}-\bg_{\mu\nu}\right)\,h^{\mu\nu}
\ .
\ee
\label{ProjGrav}
\end{subequations}
Off-shell, there are five degrees of freedom coming from $h_{\mu\nu}$, three from the vector and one from each of the two scalars. 
This matches the number of entries of a $4\times 4$ symmetric matrix.
\section{Enter the inflation: approximate de Sitter background}
The most recent data on CMB~\cite{CMB} give an estimate on the value of the parameter 
$\alpha/\lp^2\sim 10^8$ during inflation. Thus, in an high curvature regime, such as the one would expect for the early Universe, the $R^2$ term dominates in~\eqref{StaroAction}.
We can start by considering an approximated background, \emph{i.e.} the solution of the $R^2$ theory, which is naively obtained by imposing $\alpha\gg1$.
The equations of motion then read,
\be
\left(2 R_{\mu\nu}-\frac{1}{2}g_{\mu\nu} R\right) R
+2(g_{\mu\nu}\Box-\nabla_\mu\nabla_\nu) R
\,=\,
0
\ .
\label{FEdS}
\ee
\par
One can easily infer that the latter admit the de Sitter space as a solution, that expressed in static coordinates reads
\be
\d s^2
\!\!&= &\!\!
\bg_{\mu\nu} \d x^\mu \d x^\nu
\notag
\\
\!\!&=&\!\!
-\left(1-\frac{r^2}{L^2}\right)\d t^2 + \left(1-\frac{r^2}{L^2}\right)^{-1} \d r^2 
+r^2 \d \Omega^2 
\ ,
\label{DeSitterSpace}
\ee
where $L$ is known as the curvature radius and $\d\Omega^2=\d \theta^2+\sin^2 \theta \d \phi^2$.
Then, from Eq.~\eqref{DeSitterSpace} one finds that
$
\bar R_{\mu\nu}
=
\Lambda \bg_{\mu\nu}
$
and $\bar R=4\Lambda$, with $\Lambda=3/L^2$ denoting the cosmological constant.
Perturbing Eq.~\eqref{FEdS} over the de Sitter background, we get
\be 
2\bar R\, \delta R_{\mu\nu}+2\bar R_{\mu\nu}\,\delta R-\bg_{\mu\nu}\bar R\,\delta R
-\frac{1}{2}h_{\mu\nu}\bar R^2+2\left(\bg_{\mu\nu}\bar\Box-\babla_\mu\babla_\nu\right)\,\delta R
\,=\,
0
\ ,
\ee
which yields
\be
-4\Lambda\left(\bar\Box h_{\mu\nu}-\babla^\lambda\babla_\mu h_{\lambda\nu}
-\babla^\lambda\babla_\nu h_{\mu\lambda}+\frac{1}{2}\babla_\mu\babla_\nu h+\frac{1}{2}\bg_{\mu\nu}\babla^\lambda\babla^\rho h_{\lambda\rho}\right)
&&
\notag
\\
+2\Lambda^2\left(\bg_{\mu\nu}h-4h_{\mu\nu}\right)-2\left(\bg_{\mu\nu}\bar\Box-\babla_\mu\babla_\nu\right)
\left(\bg_{\lambda\rho}\bar\Box-\babla_\lambda\babla_\rho\right)h^{\lambda\rho}
\!\!&=&\!\!
0
\ .
\ee
It is convenient to commute two covariant derivatives,
\be
\babla^\lambda\babla_\nu h_{\mu\lambda}
\!\!&=&\!\!
\babla_\nu\babla^\lambda h_{\mu\lambda}+\bar R_{\lambda\nu}h^\lambda_{\,\,\mu}
-\bar R^\rho_{\,\,\mu\lambda\nu}h^\lambda_{\,\,\rho}
\notag
\\
\!\!&=&\!\!
\babla_\nu\babla^\lambda h_{\mu\lambda}+\frac{\Lambda}{3}\left(4h_{\mu\nu}-\bg_{\mu\nu}h\right)
\ ,
\label{Commutation}
\ee
so that the field Equations become
\be
-4\Lambda\left(\bar\Box h_{\mu\nu}-\babla_\mu\babla^\lambda h_{\lambda\nu}
-\babla_\nu\babla^\lambda h_{\mu\lambda}+\frac{1}{2}\babla_\mu\babla_\nu h+\frac{1}{2}\bg_{\mu\nu}\babla^\lambda\babla^\rho h_{\lambda\rho}\right)
&&
\notag
\\
+\frac{2\Lambda^2}{3}\left(4h_{\mu\nu}-\bg_{\mu\nu}h\right)-2\left(\bg_{\mu\nu}\bar\Box-\babla_\mu\babla_\nu\right)
\left(\bg_{\lambda\rho}\bar\Box-\babla_\lambda\babla_\rho\right)h^{\lambda\rho}
\!\!&=&\!\!
0
\label{FEcommuted}
\ .
\ee
\par
Before inserting the projections~\eqref{ProjGrav}, it is appropriate to reduce Eq.~\eqref{FEcommuted}
in Laplacian form by choosing a convenient gauge fixing. This allows to compute the right number
of propagating degrees of freedom. According to the analysis in Section~\ref{sec:GF}, we choose the condition
\be
\babla^\mu h_{\mu\nu}
\,=\,
\frac{1}{4}\babla_\nu h
\ ,
\label{GFdS}
\ee
which resembles the usual de Donder gauge fixing on a Minkowskian background. Moreover,
we can use the latter to fix $h'_{\mu\nu}$ in~\eqref{hGauge} to be $h^\bot_{\mu\nu}$, which satisfies the gauge fixing condition, thanks to~\eqref{TransvTraceless}.
Thus,
\be
-4\Lambda\left(\bar\Box h_{\mu\nu}+\frac{1}{8}\bg_{\mu\nu}\bar\Box h\right)
+\frac{2\Lambda^2}{3}\left(4h_{\mu\nu}-\bg_{\mu\nu}h\right)-\frac{3}{2}\left(\bg_{\mu\nu}\bar\Box-\babla_\mu\babla_\nu\right)\bar\Box h
\!\!&=&\!\!
0
\ .
\ee
We can now insert the projections introduced in~\eqref{ProjGrav}. As we anticipated,
the EOM propagate a tensor and a scalar mode~\cite{lust}, as the equations which
do not vanish identically are
\be
\bar\Box \left(\bar \Box+\frac{4\Lambda}{3} \right) \varphi
\!\!&=&\!\!
0
\ ,
\\
\left(\bar \Box -\frac{2\Lambda}{3}\right)h^\bot_{\mu\nu}
\!\!&=&\!\!
0
\ ,
\label{dS_STF}
\ee
which correspond to Eqs.~\eqref{TraceEOM} and~\eqref{EOMreduced}, respectively.
The tensor mode propagates freely and it possesses an effective mass, proportional to the 
inverse Hubble parameter $H^2=\frac{\Lambda}{3}$, which is due to the fact that any wavelength cannot exceed
the radius of the cosmological horizon. Therefore, the energy spectrum is gapped and
\be
m_{\rm eff}^2
\,=\,
\frac{2\hbar^2\Lambda}{3}
\,=\,
2\hbar^2H^2
\ .
\ee
Concerning the scalar field, it seems to have two kind of dispersion relations: a massless and a tachyonic one. This latter is due to the particle creation effect in a de Sitter
background and is directly related to its unstable behavior. An eternal de Sitter geometry
is indeed non-physical.
\section{True inflationary phase}
Approximating the inflationary metric with de Sitter gives some glances, but not the full
picture. In order to get it, we have to consider the full field Equation~\eqref{EOM}. Anyway,
since we saw that it splits in a free EOM for any of the propagating degrees of freedom,
introduced in~\eqref{ProjGrav}, we will consider Eqs.~\eqref{TraceEOM} and~\eqref{EOMreduced}, directly.
\par
First, we consider the scalar field $\varphi$. The linearisation of~\eqref{TraceEOM} reads
\be
6\alpha \left(\bar \Box \delta R
-\bg^{\mu\nu}\delta \Gamma^\lambda_{\mu\nu}\babla_\lambda \bar R
-h_{\mu\nu}\babla^\mu\babla^\nu \bar R\right)
-\delta R
\,=\,
0
\ ,
\ee
which is, after some straightforward algebra
\begin{multline}
\left(1-6\alpha\bar \Box\right)\left(\bg^{\mu\nu}\bar \Box -\babla^\mu\babla^\nu\right)h_{\mu\nu}
-6\alpha\left[ \bar \Box\left(\bar R^{\mu\nu}h_{\mu\nu}\right)+
\left(\babla^\mu h_{\mu\nu}-\frac{1}{2}\babla_\nu h\right)\babla^\nu \bar R
+h_{\mu\nu}\babla^\mu\babla^\nu \bar R
\right]
\\
+\bar R^{\mu\nu}h_{\mu\nu}
\,=\,
0
\ .
\end{multline}
Recalling that $\varphi$ enters through $h_{\mu\nu}\supset \frac{1}{3}\bg_{\mu\nu}\varphi$,
we obtain
\be
-6\alpha \bar \Box^2 \varphi +\left(1-2\alpha \bar R\right)\bar \Box\varphi 
-\frac{1}{3}\bar R \varphi +2\alpha\babla^\mu \bar R \babla_\mu \varphi
\,=\,
0
\ ,
\ee
where we made use of the background equation $6\alpha \bar\Box \bar R=\bar R$. 
Dividing by $-6\alpha$ and, once again, taking profit of the background equation we get
\be
\bar \Box\left[\bar \Box -\frac{(1-2\alpha\bar R)}{6\alpha}\right]\varphi
-\frac{1}{3}\babla^\mu \bar R \babla_\mu \varphi
\,=\,
0
\ .
\ee
\par
If we insert the mass scale $M$, defined in~\eqref{StaroAction}, the kinetic part of 
this Equation becomes
\be
\bar \Box\left[\bar \Box -\frac{(1-2\alpha\bar R)}{6\alpha}\right]\varphi
\,=\,
\bar \Box\left[\bar \Box -\left(\frac{M^2}{\lp^2\mpl^2}-\frac{\bar R}{3} \right)\right]\varphi
\ .
\ee
It shows that this scalar field, which has been identified with the inflaton~\cite{Casadio:2017twg}, shifts from a tachyonic behavior to an oscillatory one at the threshold ``background curvature"
\be
\bar R_{\rm crit}
\,=\,
\frac{3 M^2}{\lp^2\mpl^2}
\ .
\ee
This feature supports the claim that the Einstein-Hilbert part of the action~\eqref{StaroAction}
behaves as a perturbation, that drives the model out of the inflationary phase, and there is
no need to introduce a new particle {\it ad hoc}, nor to employ some transformation
to the Einstein frame. However, we stress out that the reader should take this result with a grain of salt.
Although this hint may look promising, it is not a definitive proof, since we are not considering (self-)interactions of $h_{\mu\nu}^\bot$ and $\varphi$, which 
are essential in determining the slow-roll parameters out of inflation through 
the analysis of a suitable potential.
\par 
\subsection{Gauge Fixing}
\label{sec:GF}
Before proceeding to the transverse, trace-free
degree of freedom $h_{\mu\nu}^\bot$, it is worthwile to 
discuss how we are going to impose the gauge fixing on the 
field equations and to generalise Eq.~\eqref{GFdS}.
Let us then consider the first order expansion of Eq.~\eqref{EOM}
\begin{multline}
\left(1+2\alpha\bar R\right)\delta R_{\mu\nu}+\left[2\alpha\bar R_{\mu\nu}-\frac{1}{2}\bg_{\mu\nu}
-\alpha\bar R \bg_{\mu\nu}+2\alpha(\bg_{\mu\nu}\bar \Box-\babla_\mu\babla_\nu)\right]\delta R
\\
+\left[2\alpha\bar \Box \bar R-\frac{\bar R}{2}\left(1+\alpha \bar R\right)\right]h_{\mu\nu}
-2\alpha \bg_{\mu\nu}\left(\babla^\lambda\babla^\rho \bar R \right)h_{\lambda\rho}
+2\alpha 
\left(
\delta \Gamma^\lambda_{\mu\nu}
-\bg_{\mu\nu}\bg^{\rho\sigma}\delta\Gamma^\lambda_{\rho\sigma}
\right)\babla_\lambda\bar R
\,=\,
0
\ ,
\end{multline}
that gives
\begin{multline}
\frac{1}{2}\left(1+2\alpha\bar R\right)\left[-\bar\Box h_{\mu\nu}+\babla_\mu\babla^\lambda h_{\lambda\nu}
+\babla_\nu\babla^\lambda h_{\mu\lambda}-\babla_\mu\babla_\nu h\right]
\\+\left[\frac{1}{2}\bg_{\mu\nu}\left(1+2\alpha\bar R\right) -2\alpha\bar R_{\mu\nu}\right]\left(\bg^{\lambda\rho}\bar \Box-\babla^\lambda\babla^\rho\right)h_{\lambda\rho}
-2\alpha(\bg_{\mu\nu}\bar\Box-\babla_\mu\babla_\nu)(\bar R^{\lambda\rho}h_{\lambda\rho})
\\
+\alpha\left[\babla_\mu h_{\lambda\nu}+\babla_\mu h_{\lambda\nu}-
\babla_\lambda h_{\mu\nu}-2\bg_{\mu\nu}\left(\babla^\rho h_{\lambda\rho}
-\frac{1}{2}\babla_\lambda h\right)\right]\babla^\lambda \bar R
\\
+\left[\left(\frac{1}{2}\bg_{\mu\nu}-2\alpha \bar R_{\mu\nu}+\alpha\bg_{\mu\nu}\bar R\right)\bar R^{\lambda\rho}h_{\lambda\rho} +\left(2\alpha\bar \Box \bar R-\frac{\bar R}{2}\left(1+\alpha \bar R\right)\right)h_{\mu\nu}-2\alpha\bg_{\mu\nu}\left(\babla^\lambda\babla^\rho \bar R \right)h_{\lambda\rho}\right]
\\
+\frac{1}{2}\left(1+2\alpha \bar R\right)\left(\bar R^\lambda_{\,\,\mu}h_{\lambda\nu}
+\bar R^\lambda_{\,\,\nu}h_{\mu\lambda}
-2\bar R^\rho_{\,\,\mu\lambda\nu}h^\lambda_{\,\,\rho}\right)
-2\alpha\left(\bg_{\mu\nu}\bar \Box-\babla_\mu\babla_\nu\right)\left(\bg^{\lambda\rho}\bar\Box-\babla^\lambda\babla^\rho\right)h_{\lambda\rho}
\,=\,
0
\ ,
\label{FieldEqTotal}
\end{multline}
unpacking $\delta R$ and $\delta \Gamma ^\lambda _{\mu \nu}$.
\par
It is convenient to simplify these equations, by imposing a gauge fixing. We can guess its form,
by making the following requirements:
\begin{itemize}
\item By pure tensorial structure the only two terms we can use are the covariant gradient
of a combination of $\bar R^{\mu\nu}h_{\mu\nu}$  and $\bar R h$, plus $h_{\mu\nu}\babla^\nu \bar R$;
\item It has to reduce to Eq.~\eqref{GFdS} for the pure $R^2$ theory;
\item The gauge fixing condition is imposed off-shell, hence it cannot contain physical degrees of freedom;
\item It must contain the combination $\babla_\nu \left(\bar R^{\lambda\rho} h_{\lambda\rho}\right)+h_{\lambda\nu}\babla^\lambda \bar R$, in order to have the desired cancelation of the gradients in the field equation~\eqref{FieldEqTotal}.
\end{itemize}
Henceforth, we get
\be
\babla^\lambda h_{\lambda\nu}-\frac{1}{2}\babla_\nu h
\,=\,
\frac{1}{\bar R}\left[\babla_\nu \left(\bar R^{\lambda\rho} h_{\lambda\rho}-\frac{1}{2}\bar R h\right)+h_{\lambda\nu}\babla^\lambda \bar R\right]
\label{GFApp}
\ .
\ee
The insertion $h_{\mu\nu}\supset \frac{1}{3}\bg_{\mu\nu}\varphi$ reduces it to
\be
\bar R \babla_\nu \varphi
\,=\,
\babla_\nu \left(\bar R \varphi\right)-\varphi \babla_\nu \bar R
\ ,
\ee
which vanishes trivially as expected. Since any gauge-fixing condition is off-shell, 
$h_{\mu\nu}^\bot$ has not to appear inside it either. Unfortunately, it is very hard to notice 
this explicitly, we do not know the exact form of the background metric $\bg_{\mu\nu}$
nor of the polarisation tensor, but we can be confident that they are such that the relation
\be
\babla_{\nu}\left(\bar R^{\lambda\rho}h_{\lambda\rho}^\bot\right)
\,=\,
-h^\bot_{\lambda\nu}\babla^\lambda \bar R
\ee
is in effect.
As a final remark, also the de Donder gauge is promptly 
recovered in the case of $\bar R=0$. To see it, one can just impose this constraint
in the field equations of the background getting $\bar R_{\mu\nu}=0$,
whose linearisation is straightforward. 
This implies the limit
\be
\underset{\bar R \to 0}{\lim}
 \left\{\frac{1}{\bar R}\left[\babla_\nu \left(\bar R^{\lambda\rho} h_{\lambda\rho}-\frac{1}{2}\bar R h\right)+h_{\lambda\nu}\babla^\lambda \bar R\right]\right\}
\,=\, 
0
\ ,
\ee
in~\eqref{GFApp}.
\subsection{Transverse traceless excitations}
Of course, we also have to check the Symmetric Trace-Free Equation~\eqref{EOMreduced}. Its linearisation reads
\begin{multline}
\left(1+2\alpha\bar R\right)\delta R_{\mu\nu}+\left(2\alpha\bar R_{\mu\nu}-\frac{1}{6}\bg_{\mu\nu}
-\alpha\bar R \bg_{\mu\nu}-2\babla_\mu\babla_\nu\right)\delta R
\\
-\frac{\bar R}{2}\left(\frac{1}{3}+\alpha \bar R\right)h_{\mu\nu}+2\alpha \delta \Gamma^\lambda_{\mu\nu}\babla_\lambda \bar R
\,=\,
0
\ .
\end{multline}
We can use Eqs.~\eqref{deltaRmn} and~\eqref{deltaR} to manipulate this latter field equation,
to get
\begin{multline}
\frac{1}{2}\left(1+2\alpha \bar R\right)\left(
-\bar \Box h_{\mu\nu}+\babla^\lambda \babla_\mu h_{\lambda\nu}
+\babla^\lambda \babla_\nu h_{\mu\lambda}-\babla_\mu\babla_\nu h
\right)
\\
+\left(\frac{1}{6}\bg_{\mu\nu}-2\alpha\bar R_{\mu\nu}+\alpha\bar R \bg_{\mu\nu}
\right)\bar R^{\lambda\rho}h_{\lambda\rho}
+\left(\frac{1}{6}\bg_{\mu\nu}-2\alpha\bar R_{\mu\nu}+\alpha\bar R \bg_{\mu\nu}
\right)\left(\bg^{\lambda\rho}\bar \Box-\babla^\lambda\babla^\rho\right)h_{\lambda\rho}
\\
2\alpha\babla_\mu\babla_\nu \left(\bar R^{\lambda\rho}h_{\lambda\rho}\right)
+\alpha\left(\babla_\mu h_{\lambda\nu}+\babla_\nu h_{\mu\lambda}
-\babla_\lambda h_{\mu\nu}\right)\babla^\lambda \bar R
-\frac{\bar R}{2}\left(\frac{1}{3}+\alpha\bar R\right)h_{\mu\nu}
\\
+2\alpha\babla_\mu\babla_\nu \left(\bg^{\lambda\rho}\bar \Box-\babla^\lambda\babla^\rho\right) h_{\lambda\rho}
\,=\,
0
\ ,
\label{STFLinear}
\end{multline}
after some cumbersome algebra (see Appendix~\ref{sec:GF}). We can make it more streamlined by means of the identification
$h_{\mu\nu}\sim h^\bot_{\mu\nu}$ (we saw it is the only component of the graviton that appears), the commutation~\eqref{Commutation} and the gauge fixing condition
\be
\babla^\lambda h_{\lambda\nu}-\frac{1}{2}\babla_\nu h
\,=\,
\frac{1}{\bar R}\left[\babla_\nu \left(\bar R^{\lambda\rho} h_{\lambda\rho}-\frac{1}{2}\bar R h\right)+h_{\lambda\nu}\babla^\lambda \bar R\right]
\ ,
\label{GaugeFixingR+R^2}
\ee
which is equivalent to
\be
\babla_\nu \left(\bar R^{\lambda\mu}h^\bot_{\lambda\mu}\right)
\,=\,
-h^\bot_{\mu\nu}\babla^\mu \bar R
\ .
\ee
In particular, it is very easy to see that the relation~\eqref{GaugeFixingR+R^2} reduces to~\eqref{GFdS} consistently
for a de Sitter background and to de Donder gauge-fixing for linearisations over a Minkowskian metric. The result is
\begin{multline}
-\frac{1}{2}\left(1+2\alpha \bar R\right)\bar \Box h^\bot_{\mu\nu}+\left(
\frac{1}{6}\bg_{\mu\nu}-2\alpha\bar R_{\mu\nu}+\alpha\bg_{\mu\nu}\bar R
\right)\bar R^{\lambda\rho}h^\bot_{\lambda\rho}-\alpha h^\bot_{\mu\lambda}\babla_\nu\babla^\lambda \bar R
-\alpha h^\bot_{\lambda\nu}\babla_\mu\babla^\lambda \bar R
\\
-\alpha \babla_\lambda h^\bot_{\mu\nu}\babla^\lambda \bar R-\frac{\bar R}{2}\left(\frac{1}{3}+\alpha\bar R\right)h^\bot_{\mu\nu}
+\frac{1}{2}(1+2\alpha \bar R)\left(\bar R^\lambda_{\,\,\mu}h_{\lambda\nu}
+\bar R^\lambda_{\,\,\nu}h_{\mu\lambda}-2\bar R^\rho_{\,\,\mu\lambda\nu}h^\bot{}^\lambda_{\,\,\rho}\right)
\,=\,
0
\ .
\label{EOMhbot}
\end{multline}
\par
The trace of this field equation, together with the relation~\eqref{h_barEOM}, yields
\be
\bar R^{\mu\nu}h^\bot_{\mu\nu}
\,=\,
0
\ ,
\ee
since
\be
0
\!\!&=&\!\!
\left(\frac{2}{3}+2\alpha \bar R\right)\bar R^{\mu\nu}h^\bot_{\mu\nu}
-2\alpha h^\bot_{\mu\nu}\babla^\mu\babla^\nu \bar R
\notag
\\
\!\!&=&\!\!
\left(\frac{2}{3}+2\alpha \bar R\right)\bar R^{\mu\nu}h^\bot_{\mu\nu}-\left(1+2\alpha \bar R\right)\bar R^{\mu\nu}h^\bot_{\mu\nu}
\notag
\\
\!\!&=&\!\!
-\frac{1}{3}\bar R^{\mu\nu}h^\bot_{\mu\nu}
\ .
\ee
The gauge fixing condition tells that this condition also involves $h^\bot_{\mu\nu}\babla^\nu \bar R=0$.
In addition to that, it is convenient to use the general decomposition of the Riemann tensor~\eqref{RiemannDecomposition}
to work out
\be
\frac{1}{2}\left(\bar R^\lambda_{\,\,\mu}h_{\lambda\nu}
+\bar R^\lambda_{\,\,\nu}h_{\mu\lambda}-2\bar R^\rho_{\,\,\mu\lambda\nu}h^\bot{}^\lambda_{\,\,\rho}\right)
\!\!&=&\!\!
\bar R^\lambda_{\,\,\mu}h^\bot_{\lambda\nu}
+\bar R^\lambda_{\,\,\nu}h^\bot_{\mu\lambda}-\frac{1}{6}\bar R h^\bot_{\mu\nu}
-\bar C^\rho_{\,\,\mu\lambda\nu}h^\bot{}^\lambda_{\,\,\rho}
\ .
\ee
The EOM reads
\begin{multline}
\bar \Box h^\bot_{\mu\nu}+\frac{2\alpha}{1+2\alpha R}\babla_\lambda h^\bot_{\mu\nu}\babla^\lambda \bar R+\bar R^\lambda_{\,\,\mu}h^\bot_{\lambda\nu}
+\bar R^\lambda_{\,\,\nu}h^\bot_{\mu\lambda}
-\frac{\bar R\left(2+5\alpha\bar R\right)}{3(1+2\alpha\bar R)}h^\bot_{\mu\nu}
-\bar C^\rho_{\,\,\mu\lambda\nu}h^\bot{}^\lambda_{\,\,\rho}
\,=\,
0
\ .
\end{multline}
\par
As desired, this EOM reduces to $\bar \Box h_{\mu\nu}^\bot=0$ for a Ricci flat solution, 
while de Sitter background yields~\eqref{dS_STF} exactly.
\section{Concluding remarks and outlook}
\label{sec:Conclusions}
In this work we clarified some statements made in~\cite{Casadio:2017twg}, about
embedding a theory of cosmological inflation in a corpuscular perspective.
To this aim, we studied the linearisation of the $f(R)=R+\alpha R^2$ theory and decomposed the
graviton on a curved background. 
We checked that the theory propagates two degrees of freedom: a spin-$2$ helicity mode
$h^\bot_{\mu\nu}$ and a scalar $\varphi$. The latter is important because its nature signals
the transition from the inflationary epoch to the reheating phase. In fact, at the beginning the 
background geometry is realised in terms of a de Sitter space, and $\varphi$
propagates tachyonic modes, which entail the instability of the exponential expansion as an
uncontrollable particle creation mechanism. On the other hand, $\varphi$ switches to a normal oscillatory
behavior when the background curvature hits a critical value $R_{\rm crit}$, and the huge
potential energy gained during the slow-roll allows the inflaton to decay, as implied by
any model of inflation. 
It is very interesting to note that the critical curvature $R_{\rm crit}$ is related to the 
coupling $\alpha$ and it emerges as a natural implication of the theory,
rather than an additional input. Furthermore, the relation with $\alpha$ makes it also measurable~\cite{CMB},
henceforth of paramount importance even on phenomenological grounds.
In summary, the corpuscular model allows to recover the standard picture of Starobinsky inflation, with the difference that the inflaton is understood as a polarisation
of the graviton and not a degree of freedom that needs a suitable field transformation in 
order to appear in the theory.

	In addition to these findings, this work paves the way for a number of future developments.
Firstly, the transition from an inflationary regime to the reheating phase is regulated by 
the emergence of a set of so-called ``slow-roll parameters"~\cite{Baumann:2009ds}. 
They vanish when the inflaton dwells on the potential plateau, while they become
$\mathcal{O}(1)$ at the end of inflation. Their formulation comes in terms of derivatives of the
potential, instead of mean curvature of the background, as our point of view suggests.
We then have to include the interactions of the propagating degrees of freedom among
each other and themselves, in such a way to check whether the analogy is correct or not.

	On a different note, it is suggestive to think about applying this algorithm to other $f(R)$-theories. They can be regarded as a rich benchmark to test the present approach. This environment has also generated a vast literature overtime, which can provide a 
useful feedback to check the results.

	Finally, we remark that this model only explains how the instabilities in the system 
drive it to an exit from inflation, and not how it is realised. One should have a very clear 
phenomenological insight of the model and a precise knowledge of the decay channels of the 
inflaton into the matter and radiation degrees of freedom. Since this task is out of reach, at least
at present time, one normally inserts non-local terms in the action~\cite{Koshelev:2016xqb,Codello:2016neo}.
Therefore, it looks very interesting to study the non-local behavior of $h^\bot_{\mu\nu}$ and $\varphi$ as a future outcome.
\section*{Acknowledgments}
The authors would like to thank L.~Berezhiani, R.~Casadio, M.~De Laurentis, G.~Dvali and
A.~Yu.~Kamenshchik for useful discussion. 
A.~Giugno is partially supported by the ERC Advanced Grant 339169 ``Selfcompletion",
as A.~Giusti is by the INFN grant FLAG. The research activity has been carried out in the framework
of activities of the National Group of Mathematical Physics (GNFM,INdAM) and the COST action
\emph{Cantata}.
\appendix
\section{Conventions and metric variations}
\setcounter{equation}{0}
\label{sec:Conv}
We follow the conventions in~\cite{Misner:1974qy}. The signature is $(-,+,+,+)$, 
ad we express fundamental constants in Planck units, that is $\hbar=\lp\mpl$ and
$G_{\rm N}=\lp/\mpl$.
\par
The Christoffel symbol
\be
\Gamma^\lambda_{\mu\nu}
\,=\,
\frac{1}{2}g^{\lambda\rho}\left(
\partial_\mu g_{\nu\rho}+\partial_\nu g_{\mu\rho}-\partial_\rho g_{\mu\nu}
\right)
\label{Christoffel}
\ ,
\ee
allows to write down the Riemann tensor
\be
R^\lambda_{\,\,\mu\rho\nu}
\,=\,
\partial_\rho \Gamma^\lambda_{\mu\nu}-\partial_\nu\Gamma^\lambda_{\mu\rho}
+\Gamma^\eta_{\mu\nu}\Gamma^\lambda_{\rho\eta}
-\Gamma^\eta_{\mu\rho}\Gamma^\lambda_{\nu\eta}
\ .
\ee
The $1\leftrightarrow 3$ contraction is the Ricci tensor $R_{\mu\nu}=R^\lambda_{\,\,\mu\lambda\nu}$, whose trace gives the curvature (Ricci) scalar $R=g^{\mu\nu}R_{\mu\nu}$.
Moreover, $R^\lambda_{\,\,\mu\rho\nu}$ can be decomposed in $4$ space-time dimensions,\be
R^\lambda_{\,\,\mu\rho\nu}
\,=\,
\frac{1}{2}\left(\delta^\lambda_{\,\,\rho}R_{\mu\nu}-g_{\mu\rho}R^\lambda_{\,\,\nu}-\delta^\lambda_{\,\,\nu}R_{\mu\rho}+g_{\mu\nu}R^\lambda_{\,\,\rho}\right)
-\frac{R}{6}\left(\delta^\lambda_{\,\,\rho}g_{\mu\nu}-\delta^\lambda_{\,\,\nu}g_{\mu\rho}
\right)+C^\lambda_{\,\,\mu\rho\nu}
\ ,
\label{RiemannDecomposition}
\ee
where $C^\lambda_{\,\,\mu\rho\nu}$ is the conformal curvature tensor.
\par
As usual, the covariant derivative of a vector $A^\mu$ is
\be
\nabla_\nu A^\mu
\,=\,
\partial_\nu A^\mu+\Gamma^\mu_{\nu\lambda} A^\lambda
\ ,
\ee
whereas it reduces to $\nabla_\mu S=\partial_\mu S$ for a scalar $S$.
The d'Alembertian operator reads $\Box=g^{\mu\nu}\nabla_\mu\nabla_\nu$ and the 
covariant derivatives commute as
\be
\nabla_\alpha \nabla_\beta A^\mu
\,=\,
\nabla_\beta \nabla_\alpha A^\mu + R^\mu_{\,\,\nu\alpha\beta} A^\nu
\label{CommuteNabla}
\ .
\ee
\par
In this Section, we vary the action~\eqref{StaroAction}, by keeping its argument a generic $f(R)$.
First, we have to remember that
\be
\frac{1}{\sqrt{-g}}\,\delta \sqrt{-g}
\,=\,
\frac{1}{2}g^{\mu\nu}\,\delta g_{\mu\nu}
\,=\,
-\frac{1}{2}g_{\mu\nu}\, \delta g^{\mu\nu}
\ ,
\ee
since $\delta g_{\mu\nu}=-g_{\mu\alpha}g_{\nu\beta}\delta g^{\alpha\beta}$.
Then, denoting $f'(R)=\delta f(R)/\delta R$, we perform the variation
\be
\frac{1}{\sqrt{-g}}\,\delta \left[\sqrt{-g} f(R)\right]
\!\!&=&\!\!
f'(R)\delta R -\frac{1}{2}f(R)g_{\mu\nu} \delta g^{\mu\nu}
\notag
\\
\!\!&=&\!\!
f'(R)g^{\mu\nu}\,\delta R_{\mu\nu}
+ \left[
f'(R)R_{\mu\nu} - \frac{1}{2}f(R)g_{\mu\nu}
\right] \, \delta g^{\mu\nu}
\ .
\ee
The variation of a Christoffel symbol, $\delta \Gamma^\lambda_{\mu\nu}$, is a tensor, unlike the symbol itself. This allows to come to the Palatini identity
\be
\delta R_{\mu\nu}
\!\!&=&\!\!
\nabla_\lambda \delta \Gamma_{\mu\nu}^\lambda-
\nabla_\nu \delta \Gamma_{\mu\lambda}^\lambda
\notag
\\
\!\!&=&\!\!
\frac{1}{2}\left(-g^{\alpha\beta}\nabla_\alpha\nabla_\beta \delta g_{\mu\nu}
+\nabla^\lambda \nabla_\mu \delta g_{\lambda\nu}
+\nabla^\lambda\nabla_\nu \delta g_{\mu\lambda}
-\Box \delta g_{\mu\nu}\right)
\label{Palatini}
\ ,
\ee
whose trace is part of the variation of the related scalar, {\emph viz.}
\be
\delta R
\!\!&=&\!\!
\delta\left(g^{\mu\nu}R_{\mu\nu}\right)
\,=\,
\delta g^{\mu\nu}R_{\mu\nu}+
\left(g_{\mu\nu}\Box -\nabla_\mu\nabla_\nu \right)\delta g^{\mu\nu}
\label{deltaRapp}
\ .
\ee
We finally have 
\be
\frac{1}{\sqrt{-g}}\,\delta \left[\sqrt{-g} f(R)\right]
\!\!&=&\!\!
\left[
f'(R) R_{\mu\nu}-\frac{1}{2}f(R)g_{\mu\nu}+
\left(g_{\mu\nu}\Box-\nabla_\mu\nabla_\nu\right) f'(R)\right]
\delta g^{\mu\nu}
\ ,
\ee
which reads
\be
\frac{1}{\sqrt{-g}}\,\delta \left[\sqrt{-g} (R+\alpha R^2)\right]
\!\!&=&\!\!
\left[
(1+2\alpha R)\, R_{\mu\nu}-\frac{1}{2}(R+\alpha R^2)g_{\mu\nu}
\right.
\notag
\\
&&\qquad 
+2\alpha
\left(g_{\mu\nu}\Box-\nabla_\mu\nabla_\nu\right) R\biggr]
\delta g^{\mu\nu}
\label{Variation}
\ee
for $f(R)=R+\alpha R^2$.
%
%

%
%


\begin{thebibliography}{99}
%
\bibitem{Casadio:2017twg}
  R.~Casadio, A.~Giugno and A.~Giusti,
  ``Corpuscular slow-roll inflation,''
  Phys.\ Rev.\ D {\bf 97} (2018) no.2,  024041
  %
  \bibitem{Guth-first}
A.H.~Guth,
Phys.\ Rev.\ D {\bf 23} (1981) 347
%
  \bibitem{starobinskyI} 
A.A.~Starobinsky,
Phys.\ Lett.\  {\bf 91B} (1980) 99;
JETP Lett.\  {\bf 30} (1979) 682
[Pisma Zh.\ Eksp.\ Teor.\ Fiz.\  {\bf 30} (1979) 719]
%
\bibitem{Linde}
A.D.~Linde,
Phys.\ Lett.\  {\bf 108B} (1982) 389
%
\bibitem{AS}
A.~Albrecht and P.J.~Steinhardt,
  Phys.\ Rev.\ Lett.\  {\bf 48} (1982) 1220
  %
\bibitem{Kofman}
  L.~Kofman, A.D.~Linde and A.A.~Starobinsky,
  ``Reheating after inflation,''
  Phys.\ Rev.\ Lett.\  {\bf 73} (1994) 3195;
  ``Towards the theory of reheating after inflation,''
  Phys.\ Rev.\ D {\bf 56} (1997) 3258
%
%
\bibitem{dvali} 
G.~Dvali and C.~Gomez,
``Quantum Compositeness of Gravity: Black Holes, AdS and Inflation,''
JCAP {\bf 1401} (2014) 023
%
\bibitem{Gia1}
G.~Dvali and C.~Gomez,
Fortsch.\ Phys.\  {\bf 61} (2013) 742;
``Black Hole's 1/N Hair,''
Phys.\ Lett.\ B {\bf 719} (2013) 419;
``Black Holes as Critical Point of Quantum Phase Transition,''
Eur.\ Phys.\ J.\ C {\bf 74} (2014) 2752
%
\bibitem{bekenstein}
J.D.~Bekenstein,
Phys.\ Rev.\ D {\bf 7} (1973) 2333
%
\bibitem{planckian}
 R.~Casadio, A.~Giugno and A.~Orlandi,
Phys.\ Rev.\ D {\bf 91} (2015) 124069;
R.~Casadio, A.~Giugno, O.~Micu and A.~Orlandi,
Entropy {\bf 17} (2015) 6893
%
\bibitem{Casadio:2017cdv}
  R.~Casadio, A.~Giugno, A.~Giusti and M.~Lenzi,
  ``Quantum corpuscular corrections to the Newtonian potential,''
  Phys.\ Rev.\ D {\bf 96} (2017) no.4,  044010
%
\bibitem{baryons}
R.~Casadio, A.~Giugno and A.~Giusti,
``Matter and gravitons in the gravitational collapse,''
Phys.\ Lett.\ B {\bf 763} (2016) 337
%
\bibitem{cko}
R.~Casadio, F.~Kuhnel and A.~Orlandi,
``Consistent Cosmic Microwave Background Spectra from Quantum Depletion,''
JCAP {\bf 1509} (2015) 002
%
\bibitem{CMB}
P.A.R.~Ade \emph{et al.}~[Planck Collaboration],
Astron.\ Astrophys.\ {\bf 594} (2016) A20 [arXiv:1502.02114 [astro-ph.CO]]
%
\bibitem{WMAP}
C.\ L.\ Bennett \emph{et al.} [WMAP Collaboration],
Astrophys.\ J.\ Suppl.\ {\bf 208} (2013) 20.
%
  \bibitem{starobinsky}
H.~Motohashi, A.A.~Starobinsky and J.~Yokoyama,
Int.\ J.\ Mod.\ Phys.\ D {\bf 20} (2011) 1347;
T.~P.~Sotiriou and V.~Faraoni,
Rev.\ Mod.\ Phys.\  {\bf 82} (2010) 451
  %
\bibitem{Capozziello:2009nq}
  S.~Capozziello, M.~De Laurentis and V.~Faraoni,
  ``A Bird's eye view of f(R)-gravity,''
  Open Astron.\ J.\  {\bf 3} (2010) 49
\bibitem{Capozziello:2011et}
  S.~Capozziello and M.~De Laurentis,
  ``Extended Theories of Gravity,''
  Phys.\ Rept.\  {\bf 509} (2011) 167
\bibitem{DeLaurentis:2015fea}
  M.~De Laurentis, M.~Paolella and S.~Capozziello,
  Phys.\ Rev.\ D {\bf 91} (2015) no.8,  083531
\bibitem{DeFelice:2010aj}
A.~De Felice and S.~Tsujikawa,
``f(R) theories,''
Living Rev.\ Rel.\  {\bf 13} (2010) 3
%
\bibitem{Cadoni:2017evg}
M.~Cadoni, R.~Casadio, A.~Giusti, W.~M\"uck and M.~Tuveri,
``Effective Fluid Description of the Dark Universe,''
Phys.\ Lett.\ B {\bf 776} (2018) 242
%
\bibitem{Cadoni:2018dnd}
  M.~Cadoni, R.~Casadio, A.~Giusti and M.~Tuveri,
  ``Emergence of a Dark Force in Corpuscular Gravity,''
  Phys.\ Rev.\ D {\bf 97} (2018) no.4,  044047
    %
\bibitem{deRham:2012kf}
  C.~de Rham and S.~Renaux-Petel,
  ``Massive Gravity on de Sitter and Unique Candidate for Partially Massless Gravity,''
  JCAP {\bf 1301} (2013) 035
  %
\bibitem{Biswas:2016etb}
  T.~Biswas, A.~S.~Koshelev and A.~Mazumdar,
  ``Gravitational theories with stable (anti-)de Sitter backgrounds,''
  Fundam.\ Theor.\ Phys.\  {\bf 183} (2016) 97
%
\bibitem{Straumann}
%
N.~Straumann,
``General Relativity",
Graduate Texts in Physics, Springer 2013, 735p
  %
\bibitem{Odintsov:1991nd}
  S.~D.~Odintsov and I.~L.~Shapiro,
  ``General relativity as the low-energy limit in higher derivative quantum gravity,''
  Class.\ Quant.\ Grav.\  {\bf 9} (1992) 873
   [Theor.\ Math.\ Phys.\  {\bf 90} (1992) 319]
   [Teor.\ Mat.\ Fiz.\  {\bf 90} (1992) 469]
\bibitem{Donoghue:2018izj}
  J.~F.~Donoghue and G.~Menezes,
  ``Gauge Assisted Quadratic Gravity: A Framework for UV Complete Quantum Gravity,''
  arXiv:1804.04980 [hep-th]
  %
\bibitem{lust}
L.~Alvarez-Gaume, A.~Kehagias, C.~Kounnas, D.~L{\"u}st and A.~Riotto,
Fortsch.\ Phys.\  {\bf 64} (2016) 176
  %
\bibitem{Baumann:2009ds}
D.~Baumann,
``Inflation,''
arXiv:0907.5424 [hep-th]
%
\bibitem{Koshelev:2016xqb}
  A.~S.~Koshelev, L.~Modesto, L.~Rachwal and A.~A.~Starobinsky,
  ``Occurrence of exact $R^2$ inflation in non-local UV-complete gravity,''
  JHEP {\bf 1611} (2016) 067
  %
\bibitem{Codello:2016neo}
  A.~Codello and R.~K.~Jain,
  ``A Unified Universe,''
  Eur.\ Phys.\ J.\ C {\bf 78} (2018) no.5,  357
  %
  %
\bibitem{Misner:1974qy}
  C.~W.~Misner, K.~S.~Thorne and J.~A.~Wheeler,
 ``Gravitation'',
  San Francisco 1973, 1279p


\end{thebibliography}
\end{document}